\documentclass[prl,twocolumn,amssymb,floatfix,amsmath,showpacs,superscriptaddress]{revtex4}
\usepackage{epsfig}

\usepackage{amssymb}
\usepackage{amsmath}
\usepackage{amsfonts}
\usepackage{bbold}
\usepackage{bm}
\usepackage{graphicx}
\usepackage{braket}
\usepackage{color}

\newcommand{\bq}{{\bf q}}
\newcommand{\br}{{\bf r}}

\begin{document}
\title{Incommensurate spin-density-wave antiferromagnetism in CeRu$_2$Al$_2$B}

\author{A. Bhattacharyya}
\email{amitava.bhattacharyya@stfc.ac.uk}
\affiliation{ISIS Facility, Rutherford Appleton Laboratory, Chilton, Didcot, Oxon, OX11 0QX, United Kingdom} 
\affiliation{Highly Correlated Matter Research Group, Physics Department, University of Johannesburg, Auckland Park 2006, South Africa}
\author{D. D. Khalyavin}
\email{dmitry.khalyavin@stfc.ac.uk}
\affiliation{ISIS Facility, Rutherford Appleton Laboratory, Chilton, Didcot, Oxon, OX11 0QX, United Kingdom} 
\author{F. Kr\"uger}
\affiliation{ISIS Facility, Rutherford Appleton Laboratory, Chilton, Didcot, Oxon, OX11 0QX, United Kingdom} 
\affiliation{London Centre for Nanotechnology, University College London, Gordon St., London, WC1H 0AH, United Kingdom}
\author{D. T. Adroja} 
\email{devashibhai.adroja@stfc.ac.uk}
\affiliation{ISIS Facility, Rutherford Appleton Laboratory, Chilton, Didcot, Oxon, OX11 0QX, United Kingdom} 
\affiliation{Highly Correlated Matter Research Group, Physics Department, University of Johannesburg, Auckland Park 2006, South Africa}
\author{A. M. Strydom} 
\affiliation{Highly Correlated Matter Research Group, Physics Department, University of Johannesburg, Auckland Park 2006, South Africa}
\affiliation{ Max Planck Institute CPfS, N\"othnitzerstr 40, 01187 Dresden, Germany}
\author{W. A. Kockelmann}
\affiliation{ISIS Facility, Rutherford Appleton Laboratory, Chilton, Didcot, Oxon, OX11 0QX, United Kingdom} 
\author{A. D . Hillier}
\affiliation{ISIS Facility, Rutherford Appleton Laboratory, Chilton, Didcot, Oxon, OX11 0QX, United Kingdom} 

\date{\today} 

\begin{abstract}
The newly discovered Ising-type ferromagnet CeRu$_2$Al$_2$B exhibits an additional phase transition at $T_\textrm{N}$ = 14.2 K before entering the ferromagnetic ground state at $T_\textrm{C}$ = 12.8 K. We clarify the nature of this transition through high resolution neutron diffraction measurements. The data reveal the presence of a longitudinal incommensurate spin-density wave (SDW) in the temperature range of $T_\textrm{C}<T<T_\textrm{N}$. The propagation vector $\bq \sim (0,0,0.148)$ is nearly temperature independent in this region and  discontinuously locks into $\bq=0$ at $T_\textrm{C}$. Mean-field calculations of an effective Ising model indicate that the modulated SDW phase is stabilized by a strong competition between ferromagnetic and antiferromagnetic exchange interactions. This makes CeRu$_2$Al$_2$B a particularly attractive model system to study the global phase diagram of ferromagnetic heavy-fermion metals under influence of magnetic frustration.

\end{abstract}


\maketitle


Heavy Fermion intermetallic compounds display a rich variety of emergent physical properties, including quantum criticality (QC), unconventional superconductivity, Kondo insulating and non-Fermi liquid states resulting from a competition between Ruderman-Kittel-Kasuya-Yosida (RKKY) exchange interaction and Kondo effect \cite{Degiorgi,Wachter,Amato}. If the RKKY interaction prevails, the system orders magnetically. Contrary, if the Kondo screening dominates, theory suggests a non-magnetic ground state where hybridization between the localized $f$ electrons and the conduction carriers opens a gap at the Fermi energy \cite{Doniach,Millis,Georges,Fulde,Hewson,Coleman}. Magnetic frustration has been recently signified as an additional dimension in the global phase diagram of heavy-Fermion metals, that can tune the degree of local-moment quantum fluctuations \cite{Si,Coleman_Nevidomskyy,QSi,Kim,Fritsch,Mun}.

Ce-based intermetallics represent an important class of heavy-fermion systems, providing a playground to study quantum criticality and the competition between RKKY and Kondo interactions. Most of them exhibit antiferromagnetic (AFM) ordering with only few examples of a ferromagnetic (FM) ground state. On the other hand, tuning of the FM transition close to the quantum critical point is of particular interest after discovery of unconventional superconductivity in some U-based compounds such as UGe$_2$~\cite{saxena}, URhGe~\cite{aoki}, UCoGe~\cite{huy} and UIr~\cite{akazawa}. 

A series of FM compounds with Ising-type anisotropy has been recently reported in the tetragonal quaternary system CeRu$_2$X$_2$M (X=Al, Ga and M=B, C) \cite{Baumbach,Matsuoka,Baumbach2,Baumbach3,Sakai}. The transition temperature to the FM state, $T_\textrm{C}$, varies from 17.2 K for CeRu$_2$Ga$_2$C down to 12.8 K for CeRu$_2$Al$_2$B \cite{Baumbach3}. In addition, the latter composition exhibits another transition at $T_\textrm{N} \sim 14.2$ K $>T_\textrm{C}$ whose nature has not been clarified so far. Based on magnetization and specific heat measurements, it is commonly believed that the transition at $T_\textrm{N}$ is magnetic but no magnetic structure determination in the temperature range of $T_\textrm{C}<T<T_\textrm{N}$ has been reported \cite{Baumbach}. Macroscopic measurements in an applied magnetic field revealed a rich ($T$-$H$) phase diagram with several stable phases. Interpretation of this behaviour also requires a precise determination of the magnetic structures in zero field. In addition, CeRu$_2$Al$_2$B is intensively studied in the context of a possible quantum phase transition tuned by chemical substitution or applied pressure \cite{Baumbach,Baumbach3,Matsuoka2}. As discussed by Baumbach \emph{et al.} \cite{Baumbach} this tuning would be especially attractive in the case of frustrated exchange interactions promoting quantum fluctuations.

In this Rapid Communication, we study the nature of the transition at $T_\textrm{N}$ by neutron powder diffraction combined with mean-field calculations of an effective Ising model. Our study reveals the onset of incommensurate SDW order  at $T_\textrm{N}$ which then discontinuously turns into the FM state at $T_\textrm{C}$. The obtained theoretical phase diagram indicates that the presence of the modulated SDW phase implies a significant competition between the nearest and next nearest neighbor interactions along the $c$-axis. This result shows that in spite of the fact that the tetragonal structure of CeRu$_2$Al$_2$B does not impose a geometrical frustration, the exchange interactions are strongly frustrated. This provides an additional degree of freedom that allows efficient tuning of the magnetic ground state in the CeRu$_2$Al$_2$B.  

Polycrystalline samples of CeRu$_2$Al$_2$B (and the non-magnetic analog LaRu$_2$Al$_2$B) were prepared by arc-melting of the constituent elements (Ce : wt.- 99.9\%, Ru : wt.- 99.9\%, Al : wt.- 99.9\%, $^{11}$B : wt.- 99.5\%) in an argon atmosphere on a water cooled copper hearth. After being  flipped and remelted several times, the boules were wrapped in tantalum foil and annealed at 1173 K for 14 days under a dynamic vacuum of  10$^{-6}$ Torr. The neutron powder diffraction data were collected on the WISH time-of-flight diffractometer at ISIS  \cite{Chapon}. The sample ($\sim $ 2 g) was loaded into a cylindrical 6 mm vanadium can and measured on warming between 1.5 K and 17 K using an Oxford Instrument cryostat. The crystal and magnetic structure Rietveld refinements were performed using the FullProf program \cite{fullprof} against the data measured in detector banks at average $2\theta $ values of $58^o$, $90^o$, $122^o$, and $154^o$, each covering $32^o$ of the scattering plane. The refinement procedure was assisted with group theoretical calculations  performed with the ISOTROPY \cite{isotropy} and ISODISTORT \cite{isodistort} software. Magnetic susceptibility and heat capacity measurements were done using Quantum Design magnetic properties (MPMS) and physical properties (PPMS) measurement systems, respectively.

\begin{figure}[t]
\centering
\includegraphics[width = \linewidth]{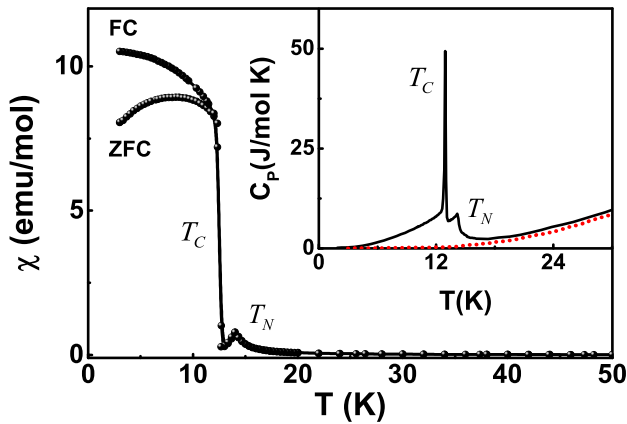}
\caption  {(Color online) Temperature dependence of dc magnetic susceptibility ($\chi$ = $M/H$) of CeRu$_2$Al$_2$B measured in zero field (ZFC) and field cooled (FC) conditions in the presence of an applied magnetic field of 100 Oe. The inset shows specific heat as a function of temperature for CeRu$_2$Al$_2$B (solid line) and its nonmagnetic analog LaRu$_2$Al$_2$B (dotted line) used to subtract the phonon contribution.}
\label{fig:F1}
\end{figure}

Fig.~\ref{fig:F1} (a) shows the magnetic susceptibility of CeRu$_2$Al$_2$B as a function of temperature collected in the magnetic field $H$ = 100 Oe. The data indicates two successive phase transitions at $T_\textrm{N}$ = 14.2 K and $T_\textrm{C}$ = 12.8 K in a good agreement with previous measurements \cite{Baumbach,Matsuoka2}. Above 50 K, the susceptibility exhibits Curie-Weiss behavior with a positive paramagnetic Curie-Weiss temperature $\theta_p$ = 10 K and an effective magnetic moment 2.35 $\mu_\textrm{B}$ close to the value 2.54 $\mu_\textrm{B}$ for a free Ce$^{3+}$-ion. The specific heat measurements also support the presence of the two transitions at $T_\textrm{N}$ and $T_\textrm{C}$ (Fig. \ref{fig:F1}, inset) and the extracted magnetic contribution of the $4f$ electrons to the entropy of the system, $S_{4f}(T)$, is practically identical to those reported by Baumbach \emph{et al.} \cite{Baumbach}.

\begin{figure}[t]
\centering
\includegraphics[scale=0.5]{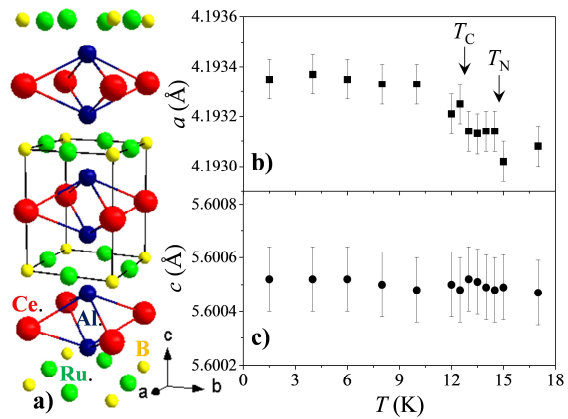}
\caption {(Color online) Schematic representation of the tetragonal crystal structure of CeRu$_2$Al$_2$B with atoms in the Wyckoff positions; Ce-$1b(0,0,1/2)$, Ru-$2f(0,1/2,0)$, Al-$2h(1/2,1/2,2563(8))$ and B-$1a(0,0,0)$ of the $P4/mmm$ space group (a). Temperature dependence of the unit cell parameters (b) and (c).}
\label{fig:F2}
\end{figure}


To explore the nature of the phase transitions at $T_\textrm{N}$ and $T_\textrm{C}$, neutron powder diffraction measurements were carried out in the temperature range of 1.5 K $<T<$ 17 K. The high temperature pattern at $T$ = 17 K was satisfactorily fitted using the tetragonal structural model  proposed by Zaikina \emph{et al.} \cite{Zaikina} for LaRu$_2$Al$_2$B [see Fig. \ref{fig:F2} (a)]. The model adopts the $P4/mmm$ space group with Ce atoms in the 1$b$(0,0,1/2) Wyckoff position, forming a square lattice and coordinated by 16-vertex cage composed of eight Ru and eight Al atoms. Below the transition at $T_\textrm{N}$, a set of new reflections at a low-Q region of the diffraction patterns appears, indicating the onset of a long range magnetic ordering. The most symmetric propagation vector which allows to account for all the observed magnetic reflections belongs to the $\Lambda$-line of symmetry $\bq = (0,0, q_c)$ with $q_c= 0.148(3)$ at $T$ = 13 K. The decomposition of the magnetic representation on the Ce $1b$ Wyckoff position consists of the two time-odd irreducible representations $m\Lambda{_4}{\oplus}m\Lambda_5$ of the $P4/mmm1'$ space group associated with the $\bq = (0,0, q_c)$ propagation vector. They transform out-of-plane and in-plane magnetic configurations respectively. There are seven symmetry distinct kernel/epikernel order parameter directions in the four-dimensional $m\Lambda_5$ representation space and only one in $m\Lambda_4$~\cite{isotropy,isodistort}.

\begin{figure}[t]
\centering
\includegraphics[scale=0.6]{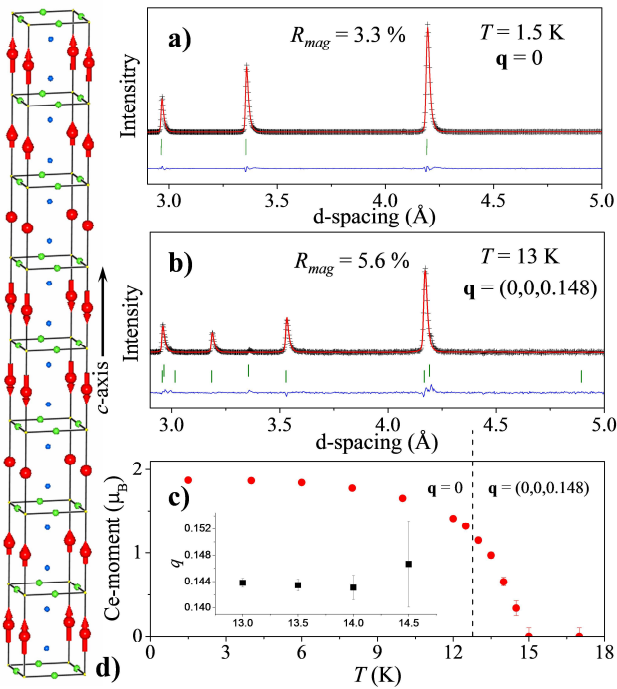}
\caption {(Color online) Rietveld refinements of the magnetic intensity of the CeRu$_2$Al$_2$B composition obtained as a difference between the diffraction patterns collected in the ordered ($T=1.5$ K and 13 K) and paramagnetic ($T=17$ K) phases. The circle symbols (red) and solid line represent the experimental and calculated intensities, respectively, and the line below (blue) is the difference between them. Tick marks indicate the positions of Bragg peaks for the magnetic scattering with the $\bq  =  0$ (a)  and $\bq = (0,0,0.148)$ (b) propagation vectors. (c) Magnetic order parameter as a function of temperature. Inset shows the temperature dependence of the incommensurate component $q_c$ of the magnetic propagation vector, $\bq = (0,0,q_c)$. (d) Incommensurate longitudinal spin-density wave representing the magnetic structure of CeRu$_2$Al$_2$B in the temperature range of $T_\textrm{C}<T<T_N$. }
\label{fig:F3}
\end{figure}

All of them were tested in the quantitative refinements using the magnetic intensities. Assuming irreducible character of the magnetic order parameter, constrained by the continuous nature of the transition at $T_\textrm{N}$ \cite{Baumbach}, the best refinement quality [Fig. \ref{fig:F3} (b)] was obtained using the $m\Lambda_4$ representation which implies a longitudinal SDW with the Ce moments being along the $c$-axis as shown in Fig. \ref{fig:F3} (d). The corresponding magnetic super space group is $P4/mmm1'$(0,0, $q_c$)00$sss$ which indicates that the phase transition at $T_\textrm{N}$ results in a loss of the translational symmetry but preserves all the rotational symmetry elements of the paramagnetic tetragonal  structure. The incommensurate component  $q_c$ of the propagation vector was found to be nearly temperature independent within the resolution of the present neutron diffraction experiment [Fig. \ref{fig:F3} (c), inset]. Below $T_\textrm{C}$, the magnetic intensity moves on top of the nuclear peaks ($\bq =  0$), consistent with the FM ground state of CeRu$_2$Al$_2$B [Fig. \ref{fig:F3} (a)]. The Ce moments stay along the $c-$axis, resulting in the $P4/mm'm'$ magnetic symmetry. The value of the magnetic moment at $T$ = 1.5 K is 1.87(3)$\mu_\textrm{B}$ and its temperature dependence is shown in Fig. \ref{fig:F3} (c). The magnetic order parameter exhibits critical behavior near $T_\textrm{N}$. The anomaly at $T_\textrm{C}$ is too small to be resolved, indicating that the first-order behaviour of the FM/SDW transition is very weak. 

Based on single crystal magnetization measurements, Matsuoka \emph{et al.} \cite{Matsuoka2} deduced the moment size 1.38 $\mu_\textrm{B}$ per Ce ion in the FM phase of CeRu$_2$Al$_2$B. This is substantially smaller than the value obtained from the present neutron diffraction data and the value expected for the $\Gamma^{(1)}_7$ doublet ground state (1.8 - 1.85 $\mu_\textrm{B}$) with the crystal field parameters evaluated from the $^{11}$B and $^{27}$Al NMR study \cite{Matsuno}. As a possible reason of the moment reduction, the authors of Ref. \cite{Matsuoka2} suggested the presence of an AFM component perpendicular to the $c$-axis. Our data do not provide any evidence of such an in-plane AFM component and are fully consistent with the simple collinear FM state. So the reason of the discrepancy between the magnetization and the neutron diffraction data remains unclear. It should be pointed out, however, that the value of the ordered moment derived in the present study is close to the moment size (1.96$\pm$0.02  $\mu_\textrm{B}$) obtained for another Ce$^{3+}$-based ferromagnet CeRu$_2$Ge$_2$ with localized $4f$-electrons \cite{Besnus} and similar ground state crystal electric field scheme \cite{Loidl}.

Let us point out that both transitions at $T_\textrm{N}$ and $T_\textrm{C}$ are not accompanied by any detectable changes in the nuclear structure. The unit cell parameters also vary very little across the transitions [Fig. 2 (b),(c)]  pointing to a weak magnetoelastic coupling. All these indicate that the scenario where the change of the magnetic structure at $T_\textrm{C}$ is structurally driven is very unlikely. 

Since the 4$f$ electrons are localized and the magnetic Ising anisotropy  large compared to the transition temperatures \cite{Matsuno}, a minimal theoretical model consists of Ising spins coupled predominantly by RKKY interactions. The RKKY exchange $J(r)$ provides a source of frustration.  It is FM on short length scales but oscillates in sign as a function of distance with a period set by the inverse Fermi momentum $1/k_\textrm{F}$. The tendency towards SDW formation can be greatly enhanced by Fermi-surface nesting or the proximity of a magnetic super-zone boundary to the Fermi surface, leading to a peak in $J(q)$ at a finite propagation vector. Such a Kohn anomaly has been argued to be responsible for the SDW formation on the border of ferromagnetism in the rare-earth metals Tb and Dy \cite{Elliot,Miwa}.  A detailed study of the RKKY interaction in CeRu$_2$Al$_2$B is beyond the scope of the present work since the electronic band structure is not known. 

\begin{figure}[t]
\centering
\includegraphics[width = \linewidth]{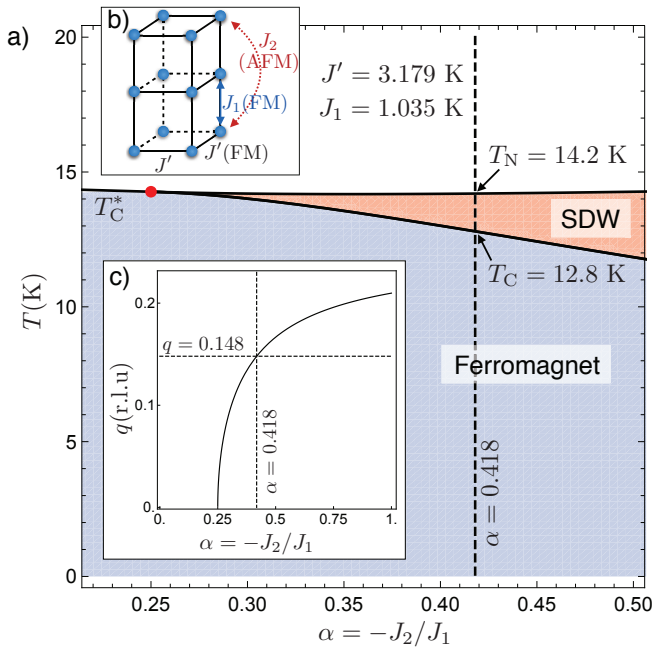}
\caption {(Color online) (a) Mean-field phase diagram of a three dimensional Ising model (exchange couplings defined in inset (b)) as a function of temperature $T$ and frustration $\alpha=-J_2/J_1$. (c) Ordering vector $\bq=(0,0,q_c)$ as a function of $\alpha$. For $J'=3.179$ K, $J_1=1.035$ K, and $J_2=-0.432$ K ($\alpha=0.418$) the theory reproduces the measured modulation vector $q_c$ and transition temperatures $T_\textrm{N}$ and $T_\textrm{C}$.}
\label{fig:F4}
\end{figure}

In the following, we use a minimal Ising model $H=-\frac 12\sum_{ij}J_{ij}\sigma_i\sigma_j$ with FM exchange interactions $J'>0$ between adjacent spins in the $a$-$b$ plane and competing FM and AFM couplings $J_1>0$ and $J_2<0$ between nearest and next nearest neigbor spins along the $c$ axis, respectively [Fig.~\ref{fig:F4}(b)]. As a function of frustration $\alpha=-J_2/J_1$, the ground state changes from a homogeneous FM to a SDW state with propagation vector $\bq=(0,0,q_c)$. We proceed to calculate the finite temperature, mean-field phase diagram. From the divergence of the magnetic susceptibility, $\chi(q)\sim [1-J(q)/T]^{-1}$, we obtain the transition temperature $T_\textrm{c}=J(q_c)$ where the propagation vector $q_c$ is determined by the maximum of $J(q)=4J'+2J_1\cos(q c)+2J_2\cos(2 q c)$. For $\alpha<1/4$, the transition is into a FM state ($q_c=0$) at 
\begin{equation}
T_\textrm{C}^*=J(0)=4J'+2J_1+2J_2.
\end{equation}
For $\alpha>1/4$, we obtain a finite propagation vector 
\begin{equation}
q_c c  =  \arccos\left(-\frac{J_1}{4J_2}  \right),
\label{Eq.qc}
\end{equation}
and a transition into a modulated SDW state at 
\begin{equation}
T_\textrm{N} =  J(q_c) = 4J'-2J_2-\frac{J_1^2}{4J_2}.
\label{Eq.TN}
\end{equation}

The FM/SDW first-order transition is obtained from a comparison of the free energies. Since the observed propagation vector $q_c= 0.148$ is small and not close to a value commensurate with the lattice we only include the first harmonic, $m(\br)=m\cos(q z)$. We can expand the free energies for small order parameters since the FM/SDW transition temperature $T_\textrm{C}=12.8$ K is close to the continuous SDW transition at $T_\textrm{N}=14.2$ K. Moreover, the discontinuity of the magnetic order parameter at $T_\textrm{C}$ is very small [Fig.~\ref{fig:F3}(c)], implying that the transition is
close to the tri-critical point at $\alpha_c=1/4$. Up to quartic order, the free energies are given by
\begin{subequations}
\begin{eqnarray}
F_\textrm{FM}(m) & = & \frac12 J(0)\left(1-\frac{J(0)}{T}\right)m^2+\frac{J^4(0)}{12T^3} m^4,\\
F_\textrm{SDW}(m,q) & = & \frac14 J(q)\left(1-\frac{J(q)}{T}\right)m^2+\frac{J^4(q)}{32T^3} m^4.\qquad
\end{eqnarray}
\label{Eq.FE}
\end{subequations}
Minimizing $F_\textrm{SDW}$ with respect to $q$ we find that in this approximation $q_c$ is independent of temperature and given by the value at the transition (\ref{Eq.qc}), consistent with the experimental observation [Fig.~\ref{fig:F3}(c)]. At low temperatures, deep in the ordered phases, the above approximations fail. In this regime it is crucial to use the 
full infinite order expressions for the free energies and to include higher harmonics of the SDW order parameter \cite{Selke}. 

Minimizing the free energies (\ref{Eq.FE}) and determining their crossing point, we obtain the temperature
\begin{equation}
T_\textrm{C} = \frac{(\sqrt{3}-\sqrt{2})T_\textrm{C}^* T_\textrm{N}}{\sqrt{3}T_\textrm{N}-\sqrt{2}T_\textrm{C}^*}
\label{Eq.FM/SDW}
\end{equation}

of the first-order FM/SDW transition. From the experimental values $q_c= 0.148$,  $T_\textrm{N}=14.2$ K, and $T_\textrm{C}=12.8$ K, and Eqs.~(\ref{Eq.qc}),(\ref{Eq.TN}), and (\ref{Eq.FM/SDW}) we obtain the effective exchange interactions $J'=3.179$ K, $J_1=1.035$ K, and $J_2=-0.432$ K. In Fig.~\ref{fig:F4}, a phase diagram as a function of temperature $T$ and frustration $\alpha=-J_2/J_1$ is shown. The experimentally relevant value $\alpha=0.418$ indicates that the frustration is significant. It is also interesting that the FM exchange in the $a$-$b$ plane is about three time stronger than along the $c$ axis. While this could partly be explained by the difference in lattice constants, $a<c$ [Fig.~\ref{fig:F2}(b,c)], and the RKKY exchange being larger at smaller distances, it probably also requires a relatively strong electronic anisotropy.

To summarize, our high resolution neutron diffraction data collected for CeRu$_2$Al$_2$B revealed the presence of a longitudinal incommensurate spin density wave in the temperature range of $T_ \textrm {C}=12.8$ K $<T<T_ \textrm {N}=14.2$ K, above the ferromagnetic phase below $T_C$. The propagation vector of the modulated phase $\bq \sim (0,0,0.148)$ is nearly temperature independent and discontinuously locks into $\bq=0$ at $T_ \textrm {C}$. The Ising model accounting for the experimentally observed behaviour implies a competition between ferromagnetic nearest $(J_1)$ and antiferromagnetic next nearest neighbor $(J_2)$ interactions along the $c$-axis. The theoretical phase diagram places CeRu$_2$Al$_2$B to the region with a substantial magnetic frustration $( -J_2/J_1 = 0.418)$.  Considering the RKKY interaction as the main source of the frustration, one can expect a strong chemical substitution/pressure dependence of the $|J_2|/J_1$ ratio.  This makes CeRu$_2$Al$_2$B  a particularly attractive model system to study the interplay between RKKY interaction, Kondo screening and quantum criticality in the global phase diagram of heavy-fermion ferromagnetic metals with an additional degree of freedom imposed by magnetic frustration.

AB thanks the FRC of UJ and ISIS-STFC for funding support. DTA and ADH would like to thank CMPC-STFC, grant number CMPC-09108, for financial support.  AMS thanks the SA-NRF (Grant 93549) and UJ Research Committee for financial support. Thanks to Prof. B.D. Rainford for providing $^{11}$B metal.

\end{document}